\documentclass{article}

\usepackage{algorithm}
\usepackage{algorithmic}
\usepackage{pdflscape}
\usepackage{longtable}
\usepackage{threeparttable}

\usepackage{natbib}

\usepackage{graphicx}
\usepackage{amssymb}

\usepackage{lineno}
\usepackage{subfigure}
\usepackage{amsmath}
\usepackage{color, colortbl}

\definecolor{Gray}{gray}{0.8}

\title{Dynamic Time Scan Forecasting}
\author{Marcelo Azevedo Costa, Leandro Brioschi Mineti \\
	Department of Production Engineering, \\Federal University of Minas Gerais, Brazil\\ \\
	\and
	Marcos Oliveira Prates,  Ramiro Ruiz C\'{a}rdenas\\
	Department of Statistics, \\Federal University of Minas Gerais, Brazil\\
	}

\date{\today}

\begin{document}

\maketitle

\begin{abstract}
The dynamic time scan forecasting method relies on the premise that the most important pattern in a time series precedes the forecasting window, i.e., the last observed values. Thus, a scan procedure is applied to identify similar patterns, or best matches, throughout the time series. As oppose to euclidean distance, or any distance function, a similarity function is dynamically estimated in order to match previous values to the last observed values. Goodness-of-fit statistics are used to find the best matches. Using the respective similarity functions, the observed values proceeding the best matches are used to create a forecasting pattern, as well as forecasting intervals. Remarkably, the proposed method outperformed statistical and machine learning approaches in a real case wind speed forecasting problem.
\end{abstract}






\section{Introduction}
\label{sec:intro}
Developing a forecasting model is not a trivial task.  Mostly, because there is no homogeneous method which achieves high accuracy in any scenario. Nevertheless, time series competitions such as the M-competitions \citep{makridakis2018statistical, makridakis2018m4}, have identified statistical and machine learning methods which have achieved consistent results. In most recent edition, the \emph{M-4 competition}, \citet{makridakis2018m4} provides evidence that combinations of mostly statistical approaches are among the most accurate methods. The second most accurate method combines statistical methods using weighted average in which machine learning is applied to estimate the weights. The authors also argue that \emph{one must accept that all forecasting approaches and individual methods have both advantages and drawbacks}. Is is also concluded that \emph{seasonal time series tend to be easier to predict}.


The motivation of this paper and the cases study is wind speed forecasting, which comprises highly complex data. Although the presence of seasonality is generally observed, the time series can behave erratically. Furthermore, wind speed forecasting is crucial for the development of alternative energy sources, mainly wind farms. However, in order to integrate wind farms with the electrical energy system both estimates of the expected energy demand and the expected generated energy are required.

Successful wind speed forecasting models are based on machine learning approaches such as neural networks, fuzzy systems, support vector machine (SVM) and hybrid systems. Mostly, because of the complexity of the time series. For instance, the use of neural networks for wind time series forecasting is quite vast \cite{bilgili2007application, cadenas2009short, li2010comparing, haque2012new}. In addition, standard statistical time series models are widely used for comparison, such as Autoregressive Moving Average (ARMA)\cite{alexiadis1998short}, Autoregressive Integrated Moving Average\cite{wang2004wind, chen2011comparison, cao2012forecasting, more2003forecasting}, Generalized Autoregressive Conditional Heteroscedastic (GARCH) \cite{guan2011short}, among others. Pre-processing of the time series is also commonly found in the literature using Spectrum Analysis (SSA) \cite{souza2012artificial}, wavelets \cite{junior2011analise, safavieh2007new, turbelin2009wavelet, bhaskar2010wind} and Fourier analysis.

An alternative class of models, known as hybrid models aim to combine machine learning models with different methods. Examples of these methods are Focused Time Delay Neural Network (FTDNN), neural networks with \emph{fuzzy} inputs \cite{hong2010hour}, finite-impulse response neural networks (FIR-NN) \cite{barbounis2006locally}, locally feedback dynamic fuzzy neural network (LF-DFNN), type recurrent fuzzy network (TRFN), neuro-fuzzy inference system (ANFIS) \cite{monfared2009new, potter2006very, pessanha2010previsao}, ARIMA-RNA \cite{shi2012evaluation}, ARIMA-SVM \cite{shi2012evaluation}, among others.

This work proposes a new model which scans the time series searching for matching windows, i.e., parts of the past time series which are similar to the last observed values from which the prediction is required. From the selected windows, the following observations are used as forecast values using similarity functions. The method named as \emph{dynamic time scan forecasting} (DTSF) is remarkably intuitive and can be used both as an exploratory tool to identify similar patterns in the time series and as an improved forecasting model. As compared to soft computing approaches, such as neural networks, the proposed method is extremely fast. Results using wind times series for a Brazilian power plant shows that the proposed method provides similar or improved prediction values as compared to statistical and machine learning models.

The proposed \emph{dynamic time scan forecasting} model was inspired by scan statistics \cite{glaz2009scan}. Scan statistics comprise a class of statistical methods which scan data streams in order to find anomalous behavior. It was introduced by Joseph Naus in 1945 \cite{naus1965distribution} and extended to epidemiological surveillance using spatial \cite{kulldorff1997spatial}, temporal and spatial-temporal data \cite{kulldorff2001prospective, kulldorff1998evaluating}. Briefly, a scanning window with fixed shape, such as a circular of a cylindrical window scans the spatial data, and a test statistic is calculated for each position of the window. The position with the largest value of the test statistic is a potential candidate for an anomalous behavior. Statistical inference is obtained using Monte Carlo simulations \cite{mooney1997monte} under the null hypothesis that the data stream does not present anomalous data. Further information about scan statistics are found in \cite{glaz2009scan}.

Similarly, the DTSF method scans a times series using a fixed window. The objective is to find previous windows in the data which are similar to the most recent observed values. Therefore, a test statistic, or a \emph{similarity statistic} is calculated for each window. In addition, a similarity function is estimated for each window. After detecting most similar windows, forecasting values are estimated using the similarity functions and the observed values which follows the selected windows. Further details about DTSF are given below.

With respect to forecasting performance, the mean squared error (MSE)\cite{wang2004wind}, mean relative error (MRE)\cite{wang2004wind}, mean absolute percentage error (MAPE) \cite{chen2011comparison, cao2012forecasting, bilgili2007application, velo2014wind, senjyu2006application}, root mean square error (RMSE) \cite{currie2014wind, velo2014wind, bechrakis2004wind} are the most common performance statistics. \citet{makridakis2018statistical} suggest using symmetric mean absolute percentage error (sMAPE) and model fitting (MF)

This work is organized as follows. Section \ref{sec:dtsf} presents the proposed DTSF method. The case study is presented in section \ref{sec:casestudy}. Results are presented in section \ref{sec:results}. Discussion and conclusion are presented in section \ref{sec:conclusion}.



\section{The dynamic time scan forecasting model}
\label{sec:dtsf}

Let $y_t$ be a time series of length $N$, $t=1,...,N$. Initially, let vector $\mathbf{y}^{[w]}$ be defined as the last $w$ observations of the time series:
\begin{equation*}
  \mathbf{y}^{[w]} = \left[ y_{N-w+1},..., y_N \right].
\end{equation*}

\begin{figure}[ht]
   \centering
   \includegraphics[scale=0.40, ]{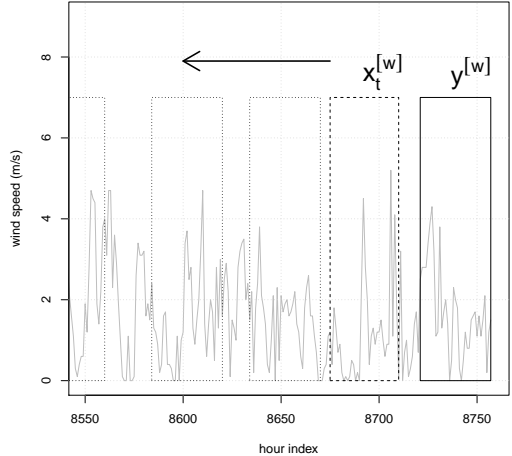}
   \caption{\label{fig:01} Illustration of the DTSF time series scan procedure.}
\end{figure}

The objective of DTSF comprises identifying patterns in the time series which are strongly correlated with vector $\mathbf{y}^{[w]}$. Thus, the set of candidate vectors can be written as:
\begin{equation*}
   \mathbf{x}_t^{[w]} = \left[ y_{t-w+1}, ..., y_{t-w} \right]
\end{equation*}
where $t=1,...,N-2 \cdot w$. The upper bound of the time sequence ($N-2 \cdot w$) guarantees that vector $\mathbf{x}_t^{[w]}$ does not overlap with vector $\mathbf{y}^{[w]}$. Figure \ref{fig:01} illustrates the DTSF method. Given the last $w$ observed values, which comprises vector $\mathbf{y}^{[w]}$, a scanning window with the same size ($\mathbf{x}_t^{[w]}$) is used to scan previous values of the time series.

The final goal of DTSF is to provide a \emph{k} steps ahead forecast of the time series, $y_{N+1}, \ldots, y_{N+k}$. To achieve this goal, the DTSF scans the time series to find the best patterns $\mathbf{x}_t^{[w]}$. The values of the time series which follow the best patterns are used as the forecast values:
\begin{equation}
  \label{eqn:03}
   y_{N+i} = f_{\mathbf{x}_t^{[w]}} (y_{t-w+i})
\end{equation}
where $f_{\mathbf{x}_t^{[w]}}$ is a function which correlates the elements of vector $\mathbf{x}_t^{[w]}$ and the elements of vector $\mathbf{y}^{[w]}$.

A first constraint can be imposed on $k$: $1\le k \le w$. This constraint guarantees that if the most correlated time series window comprises the most recent values, prior to vector $\mathbf{y}^{[w]}$, then the forecast values are a function of vector $\mathbf{y}^{[w]}$,
\begin{equation}
  \label{eqn:04}
   y_{N+i} = f_{\mathbf{x}_{N-2w}^{[w]}} (y_{N-w+i}).
\end{equation}

As observed in Equations \eqref{eqn:03} and \eqref{eqn:04}, forecast values depend on the window length $w$ and function $f_{\mathbf{x}_t^{[w]}}(\cdot)$. A first proposal for function $f_{\mathbf{x}_t^{[w]}}(\cdot)$ is a linear scaling of the elements of vector $\mathbf{x}_t^{[w]}$, i.e., a linear model. This is because previous values might supposedly be similar to the last values, except for a scale and/or offset shift. Thus, the method searches for values which might be similar to the last values, after applying a similarity function.

\begin{figure}[ht]
   \centering
   \includegraphics[scale=0.40, angle=-90]{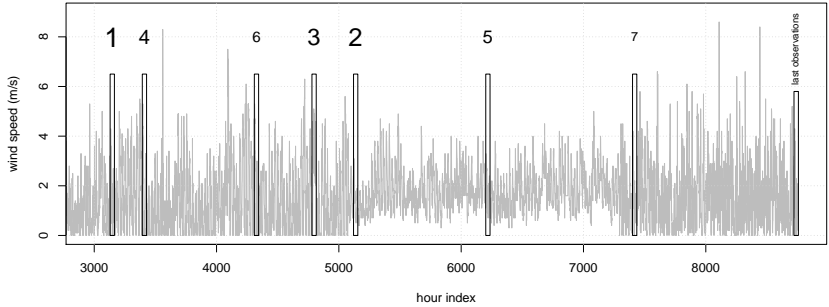}
   \caption{\label{fig:02} Example of DTSF using wind speed time series.}
\end{figure}

By assuming the similarity function as a linear model, the parameters of the model can be estimated to minimize the sum of squares between the elements of vector $\mathbf{y}^{[w]}$ and the linear equation: $\beta^{[t]}_0 + \beta^{[t]}_1 \times \mathbf{x}_t^{[w]}$. Furthermore, the similarity statistic can be defined as the linear regression coefficient of determination $R^2$ \cite{montgomery2012introduction}:
\begin{equation*}
   R^2 = 1 - \frac{ \sum_j { \left( \mathbf{y}^{[w]}_j - \hat{\mathbf{y}}^{[w]}_j \right)^2  } } { \sum_j { \left( \mathbf{y}^{[w]}_j - \bar{y}^{[w]} \right)^2  } }
\end{equation*}
where $\mathbf{y}^{[w]}_j$ is the $j$-th value of vector $\mathbf{y}^{[w]}$ and $\hat{\mathbf{y}}^{[w]}_j$ is the $j$-th predicted value using the estimated linear function. It is worth mentioning that $R^2$ is within the unit interval [0-1]. If $R^2 \rightarrow 1$ then estimated values are very close to the observed values, i.e., the past observed values located at time $t$ are similar to the last observed values after scaling and shift correction. The scanning procedure is illustrated in Figure \ref{fig:02} using a window of length 36 (hours). The past 7 windows with high similarity statistics ($R^2$) are indicated in rectangles. For each window a linear model (similarity function) was estimated.

\begin{figure}[ht]
   \centering
   \includegraphics[scale=0.40]{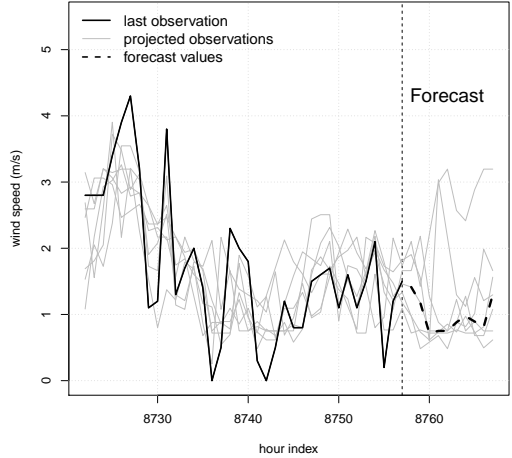}
   \caption{\label{fig:03} Forecasting estimates using the most similar windows, selected using DTSF.}
\end{figure}

Using the similarity functions, the follow up data from selected windows are used as forecast values, as illustrated in Figure \ref{fig:03}. Point estimates are generated using an aggregation function such as the median values.

The DTSF method requires three parameters: the length of the scanning window, the similarity function specification and the number of best matches, i.e., the number of similar windows found in the time series. In order to improve computational speed, linear similarity functions are preferable such as linear, quadratic and cubic linear equations. The number of best matches can be selected dynamically  using, for instance, a threshold for the similarity statistic. In this paper, a pre-defined grid of values for the length of the scanning window and the number of best matches is used. The final estimate is selected based on the minimum forecasting error in the previous day.

One important requirement of the proposed DTSF method is the availability of a large time series. The DTSF is a data driven method. Furthermore, the method can be applied to different time series using them as secondary data. For example, if wind times series are available in different weather stations across a geographical region then, in order to improve prediction in one specific location, the method can be used to select matches in neighboring weather stations. It is worth mentioning that the proposed DTSF method is available in the R package \emph{DTScanF} \cite{marcelo_costa_2019_2603008}.

\section{Case study}
\label{sec:casestudy}

The case study data comes from a wind power generation plant located in the south of the state of Bahia, in the northeastern region of Brazil. The company named \emph{Renova Energia} (\emph{Renew Energy}) owns the wind power plant and belongs to a major eletricity group named CEMIG. Wind speed (meter/second) data was collected in a measuring tower (torre Ventos do Nordeste), 78 meters high, at 10-minute intervals, from 21 November 2011 to 22 June 2016. The data was aggregated into 30-minute intervals using the mean function and has 80,398 observations.

Five days, within each weather season, were randomly selected from the final year of observations, in order to evaluate the forecasting methods. Table \ref{table:01} shows the selected dates for each weather season.

\begin{table}[ht]
   \caption{Randomly selected days for evaluating the wind speed forecasting methods.} \vspace{0.25 cm}
   \label{table:01}
   \centering
   \begin{tabular}{l l l l}
   \hline
   Winter      &  Spring      &    Summer    &    Autumn    \\ \hline
   2015-06-28  &  2015-09-30  &  2016-01-31  &  2016-04-03  \\
   2015-07-04  &  2015-10-01  &  2016-02-05  &  2016-04-05  \\
   2015-08-08  &  2015-10-26  &  2016-02-24  &  2016-04-12  \\
   2015-08-11  &  2015-12-02  &  2016-02-25  &  2016-04-13  \\
   2015-09-18  &  2015-12-06  &  2016-03-13  &  2016-05-18  \\ \hline
   \end{tabular}
\end{table}

The DTSF requires three parameters, the length of the scanning window, the similarity function and the number of best matches. These parameters were chosen based on prediction performance in the previous forecasting day. Four different similarity functions were evaluated: the linear function, $f(x) = \beta_0 + \beta_1 x$, the quadratic function, $f(x) = \beta_0 + \beta_1  x + \beta_2 x^2$, the cubic function, $f(x) = \beta_0 + \beta_1 x + \beta_2  x^2 + \beta_3 x^3$ and the polynomial function of order 4, $f(x) = \beta_0 + \beta_1 x + \beta_2  x^2 + \beta_3 x^3 + \beta_4 x^4$. In addition, both the length of the scanning window and the number of best matches were selected based on a grid of values. The investigated number of matches were 15, 25 and 50 matches. The investigated length of the scanning window were 24, 48, 72, 96 and 120 observations or 0.5, 1, 1.5 and 2 days, respectively. On total 60 different combinations of similarity functions, number of matches and length of the scanning window were investigated. The optimal parameters were chosen based on the observed mean absolute error in the previous day of forecasting.

Finally, the forecasting performance was evaluated using the $MAE=\frac{1}{k}\sum^k_{t=1} \left| y_t - \hat{y}_t \right|$,
$RMSE=\sqrt{ \frac{1}{k}\sum^k_{t=1} \left( y_t - \hat{y}_t \right)^2} $,
$sMAPE=\frac{2}{k}\sum^k_{t=1}\frac{\left| y_t - \hat{y}_t \right|}{|y_t|+|\hat{y}_t|}$
and $MF=k \times \frac{ \sum^k_{t=1} \left( y_t - \hat{y}_t \right)^2 }{ \left( \sum^k_{t=1} y_t  \right)^2 }$
statistics, where $y_t$ is the observed value, $\hat{y}_t$ is the forecast value and $k$ is the forecasting horizon.

\subsection{Forecasting methods}

Eight forecasting methods were also evaluated in order to predict the wind speed. The na\"{i}ve method replicates the observed wind speed in the previous day, i.e., the last 48 observations, as the forecast values. The ARIMA, ETS (Exponential smoothing state space model), TBATS (Exponential smoothing state space model with Box-Cox transformation, ARMA errors, Trend and Seasonal components) models were implemented using the \textbf{forecast} package \citep{robHyndman2008, robHyndman2017}. The STL + ETS (STL: Seasonal Decomposition of Time Series by Loess) model was implemented using the \textbf{stats} \citep{Rmanual} package. The NNET 1 (Feed-forward neural networks with a single hidden layer and lagged inputs for forecasting univariate time series) model was also implemented using the \textbf{forecast} package. The NNET 2 model (Extreme learning machines for time series forecasting) was implemented using the \textbf{nnfor} package \citep{Rnnfor}. The hybrid (Hybrid time series modeling) was implemented using the \textbf{forecastHybrid} package \citep{RforecastHybrid}.

\section{Results}
\label{sec:results}

\begin{figure}[H]
   \centering
   \includegraphics[scale=0.60]{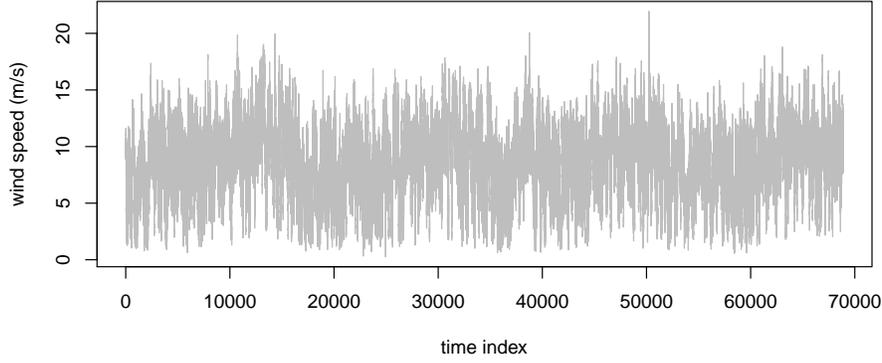}
   \caption{\label{fig:001} Time series of the wind speed (m/s).}
\end{figure}

Figure \ref{fig:001} shows the case study time series. As mentioned, it comprises approximately 70,000 observations. The average speed is 8.65 m/s. The maximum and minimum values are 21.95 m/s and 0.25 m/s, respectively. The standard deviation is 3.16 m/s.

\begin{figure}[H]
   \centering
   \subfigure[MAE]{
            \label{fig:002}
            \includegraphics[width=0.47\textwidth]{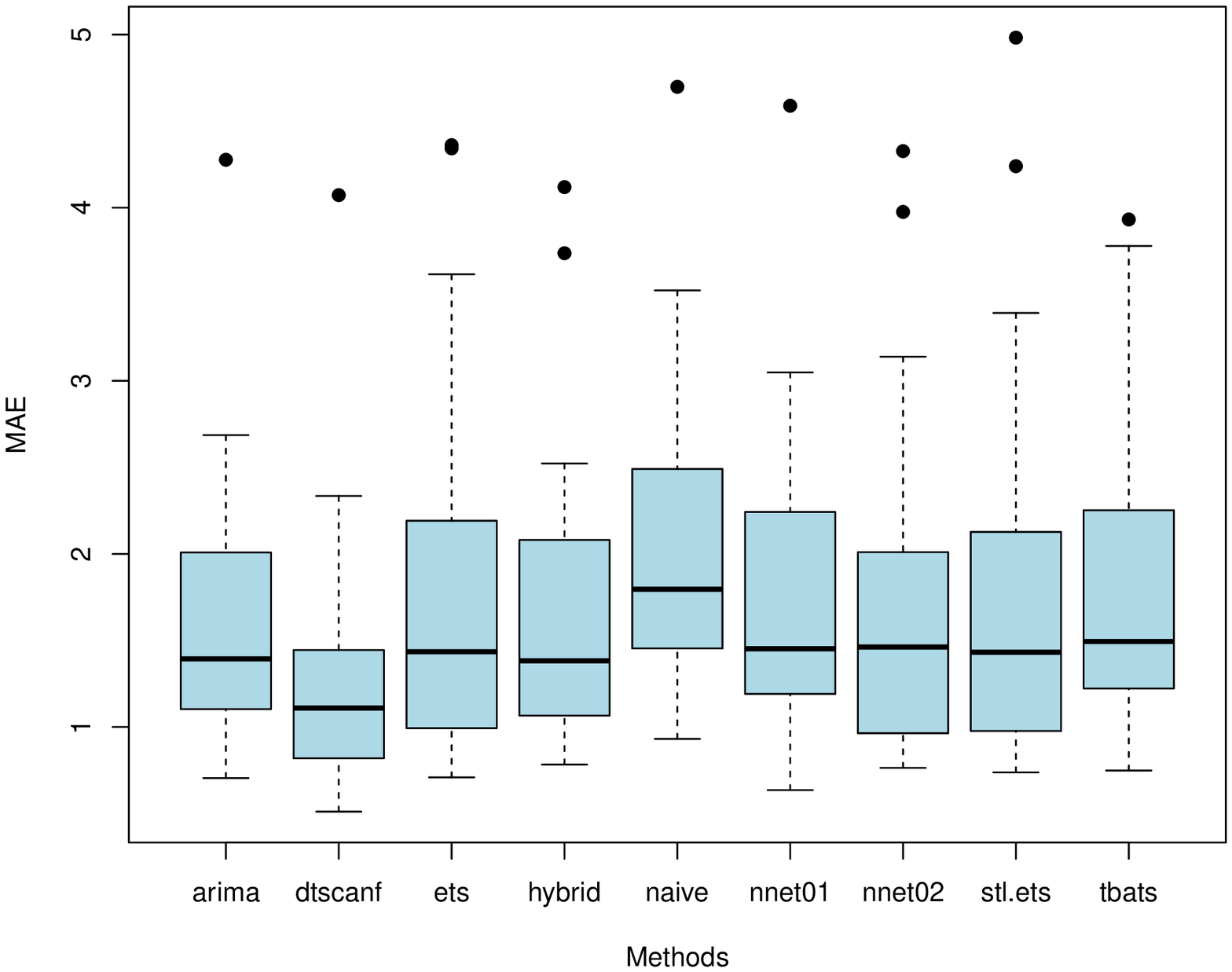}
   }
   \subfigure[RMSE]{
            \label{fig:003}
            \includegraphics[width=0.47\textwidth]{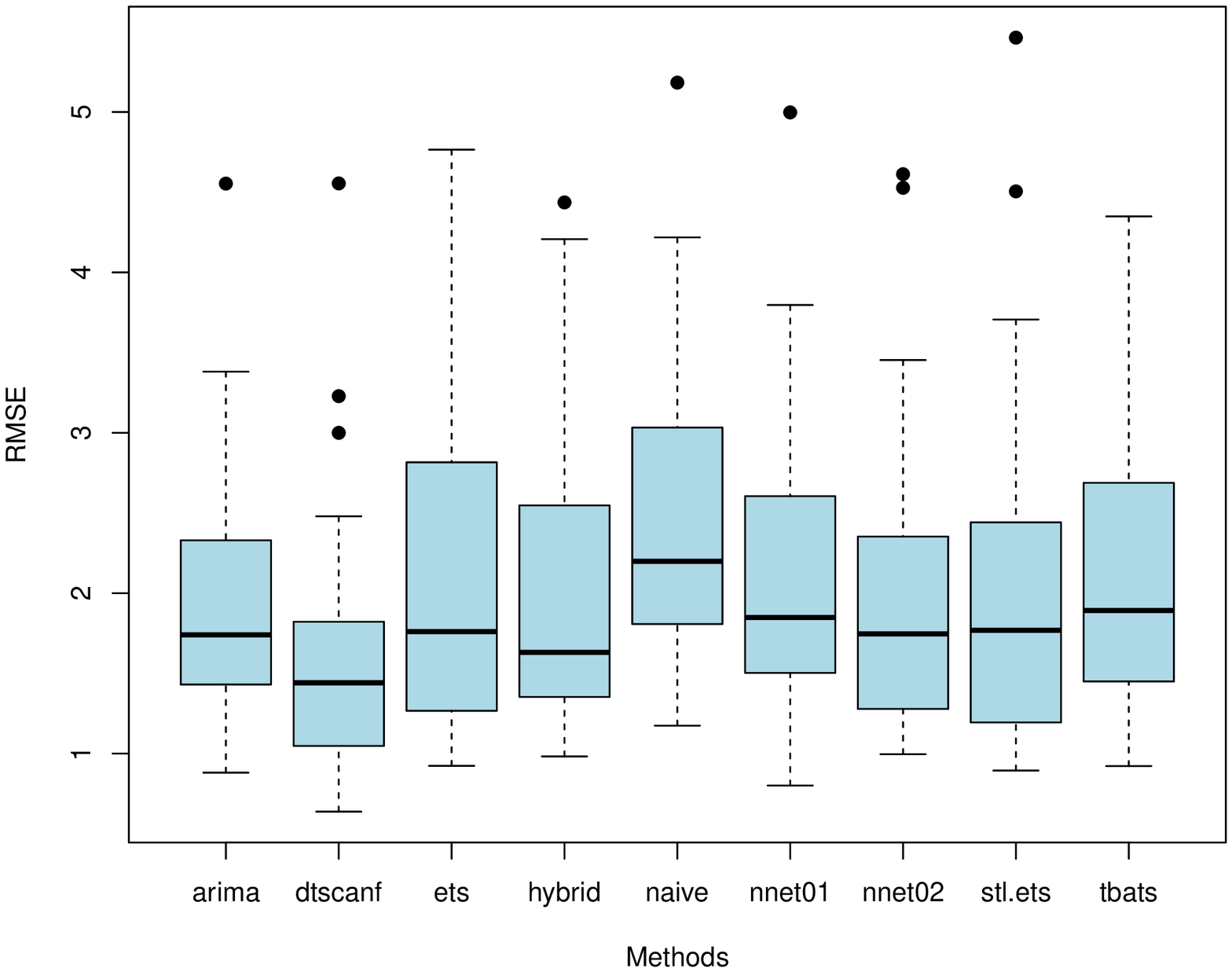}
   }
   \caption{MAE (a) and RMSE (b) statistics for the evaluated forecasting methods.}
   \label{fig:004}
\end{figure}

Table \ref{table:02} shows the error statistics and the computing time for each method. Regarding computing time, the na\"{i}ve approach is the fastest, followed by ETS, STL ETS, ARIMA and DTSF. Asterisks indicate methods which used a shorter time series in order to run. Both Hybrid and NNET.1 methods were not able to run using the total time series and, therefore, were fitted using only the last 365 days (one year) of data, or 17,520 observations prior to the forecasting day. Nevertheless, even though the Hybrid and NNET.1 methods used approximately 20\% of the total database, their computing time was large. TBATS and NNET.2 presented the largest computing times using the complete time series. Regarding MAE, RMSE sMAPE and MF error statistics, the proposed DTSF achieved the minimum values, as shown in bold type, followed by NNET.2 (MAE), Hybrid (RMSE, MF and sMAPE) and TBATS (sMAPE). Figure \ref{fig:002} shows the MAE statistics and Figure \ref{fig:003} shows the RMSE boxplots for each method. In general, the proposed DTSF presented the lower median value and the smaller quantile distance, i.e., the smaller dispersion. 

\begin{table}[H]
   \caption{Forecasting error statistics for all evaluated methods. Best results are shown in bold type.}
   \label{table:02}
   \centering
   \vspace{0.3cm}
   \begin{tabular}{l r r r r r}
   \hline
    Method  & time (sec) & MAE   & RMSE   & sMAPE   & MF    \\ \hline
    ARIMA   & 24.64      & 1.51  & 1.84   & 0.18    & 0.08  \\
    DTSF    & 31.50      & \textbf{1.11}  & \textbf{1.44}   & \textbf{0.12}    & \textbf{0.05}  \\
    ETS     & 6.43       & 1.43  & 1.76   & 0.19    & 0.10  \\
    Hybrid(*)  & 224.45     & 1.38  & 1.63   & 0.16    & 0.07  \\
    na\"{i}ve   & \textbf{0.00}       & 1.80  & 2.20   & 0.19    & 0.10  \\
    NNET.1(*)  & 737.51     & 1.45  & 1.85   & 0.17    & 0.08  \\
    NNET.2  & 2814.97    & 1.31  & 1.64   & 0.18    & 0.08  \\
    STL+ETS	& 6.82       & 1.43  & 1.78   & 0.17    & 0.08  \\
    TBATS   & 674.78     & 1.53  & 1.88   & 0.16    & 0.08  \\ \hline
   \end{tabular}
\end{table}

Table \ref{table:03} shows the average MAE statistics for each method and each  forecasting days. The two methods with minimum MAE statistics in each day are highlighted. In general, the DTSF were among the two best predictions in 14 days (70\%). In general, the remaining methods were among the best predictions in three days (15\%), on average. The ETS method, were among the best predictions in five days, followed by NNET.1 and STL+ETS for four days. Interestingly, the na\"{i}ve method achieved second best prediction in two days. 
\begin{table}[ht]
   \caption{Average MAE value for each day and forecasting method.} \vspace{0.25 cm}
   \label{table:03}
   \centering
   \footnotesize
   \begin{tabular}{c c c c c c c c c c}
   \hline
day & arima & dtsf & ets & hybrid & na\"{i}ve & nnet.1 & nnet.2 & stl+ets & tbats \\ \hline
2015-06-28 & 1.37 & \cellcolor{Gray}0.81 & 1.90 & 1.23 & 1.92 & \cellcolor{Gray}0.88 & 1.28 & 1.40 & 1.36 \\
2015-07-04 & 1.59 & 1.00 & 1.07 & 1.18 & 2.06 & \cellcolor{Gray}0.74 & \cellcolor{Gray}0.95 & 0.99 & 2.57 \\
2015-08-08 & 1.14 & 1.09 & \cellcolor{Gray}0.85 & 1.01 & 1.87 & 1.58 & 0.96 & \cellcolor{Gray}0.90 & 1.47 \\
2015-08-11 & 2.69 & \cellcolor{Gray}1.39 & 4.36 & 3.74 & \cellcolor{Gray}1.52 & 2.11 & 3.98 & 4.98 & 3.78 \\
2015-09-18 & \cellcolor{Gray}0.73 & 1.13 & \cellcolor{Gray}0.71 & 0.89 & 1.61 & 1.31 & 0.96 & 0.80 & 0.76 \\ \hline
2015-09-30 & 1.84 & 1.50 & \cellcolor{Gray}1.35 & \cellcolor{Gray}1.47 & 2.85 & 1.86 & 1.66 & 1.59 & 1.94 \\
2015-10-01 & \cellcolor{Gray}1.36 & \cellcolor{Gray}1.28 & 1.76 & 1.38 & 1.75 & 1.49 & 1.44 & 2.02 & 1.44 \\
2015-10-26 & 0.71 & \cellcolor{Gray}0.51 & 0.81 & 0.78 & 0.98 & \cellcolor{Gray}0.64 & 0.82 & 0.74 & 0.77 \\
2015-12-02 & 1.37 & \cellcolor{Gray}0.85 & 1.61 & 1.30 & 1.45 & 1.39 & 1.51 & 1.39 & \cellcolor{Gray}0.83 \\
2015-12-06 & 1.42 & \cellcolor{Gray}1.15 & \cellcolor{Gray}1.16 & 1.40 & 1.85 & 2.38 & 1.40 & 1.50 & 1.52 \\ \hline
2016-01-31 & 1.96 & \cellcolor{Gray}0.83 & 3.62 & 2.39 & 1.68 & \cellcolor{Gray}1.42 & 3.14 & 3.39 & 2.14 \\
2016-02-05 & 0.98 & 0.86 & 0.97 & 0.82 & 1.35 & 0.90 & \cellcolor{Gray}0.76 & 0.83 & \cellcolor{Gray}0.75 \\
2016-02-24 & 2.63 & \cellcolor{Gray}2.01 & 2.38 & 2.52 & 3.29 & 2.94 & 2.38 & \cellcolor{Gray}2.23 & 2.45 \\
2016-02-25 & 1.07 & \cellcolor{Gray}0.78 & 1.01 & \cellcolor{Gray}0.90 & 1.30 & 1.27 & 0.92 & 0.96 & 1.27 \\
2016-03-13 & 4.28 & \cellcolor{Gray}4.07 & 4.34 & 4.12 & 4.70 & 4.59 & 4.33 & 4.24 & \cellcolor{Gray}3.93 \\ \hline
2016-04-03 & 1.00 & \cellcolor{Gray}0.72 & \cellcolor{Gray}0.97 & 1.39 & 1.45 & 1.11 & 1.61 & 1.61 & 1.30 \\
2016-04-05 & 1.43 & \cellcolor{Gray}1.30 & 1.52 & 1.50 & 2.13 & 1.68 & 1.48 & \cellcolor{Gray}1.43 & 1.59 \\
2016-04-12 & 2.54 & 2.33 & 2.01 & 1.77 & 3.52 & 2.40 & \cellcolor{Gray}1.54 & \cellcolor{Gray}1.43 & 2.26 \\
2016-04-13 & \cellcolor{Gray}2.05 & \cellcolor{Gray}2.00 & 2.58 & 2.48 & 3.25 & 3.05 & 2.36 & 2.28 & 2.25 \\
2016-05-18 & 1.30 & \cellcolor{Gray}0.74 & 1.31 & 1.12 & \cellcolor{Gray}0.93 & 1.37 & 1.23 & 1.27 & 1.18 \\ \hline
   \end{tabular}
\end{table}

\begin{figure}[ht]
   \centering
   \includegraphics[scale=0.55]{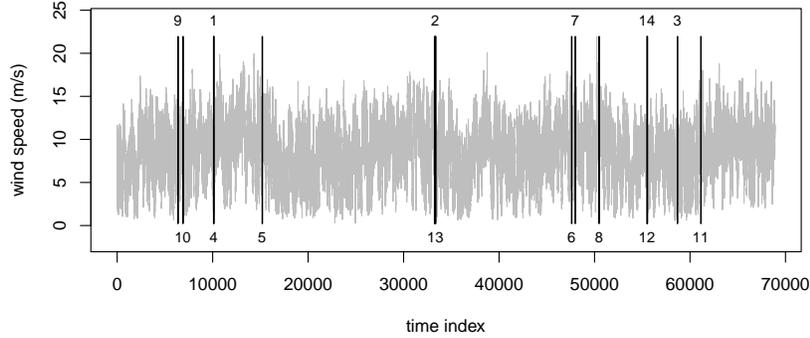}
   \caption{\label{fig:008} Time series of the wind speed (m/s) and best matches found using DTSF.}
\end{figure}

Figure \ref{fig:008} illustrates the location of the best matches for wind speed forecasting in October 26, 2015. The number of matches was estimated as 15, using the grid search procedure. It is worth noticing that the first match is located in 210.3 days before the end of the times series. The second match is located in 694.3 days before the end of the time series. The farthest match is located in 1273.35 days or 3.49 years before the end of the time series. As mentioned, after scaling and bias correction using the similarity function, the DTSF method is able to find matches which are necessarily not located close to the forecasting day.

\begin{figure}[H]
   \centering
   \subfigure[Projected values.]{
            \label{fig:005}
            \includegraphics[width=0.45\textwidth]{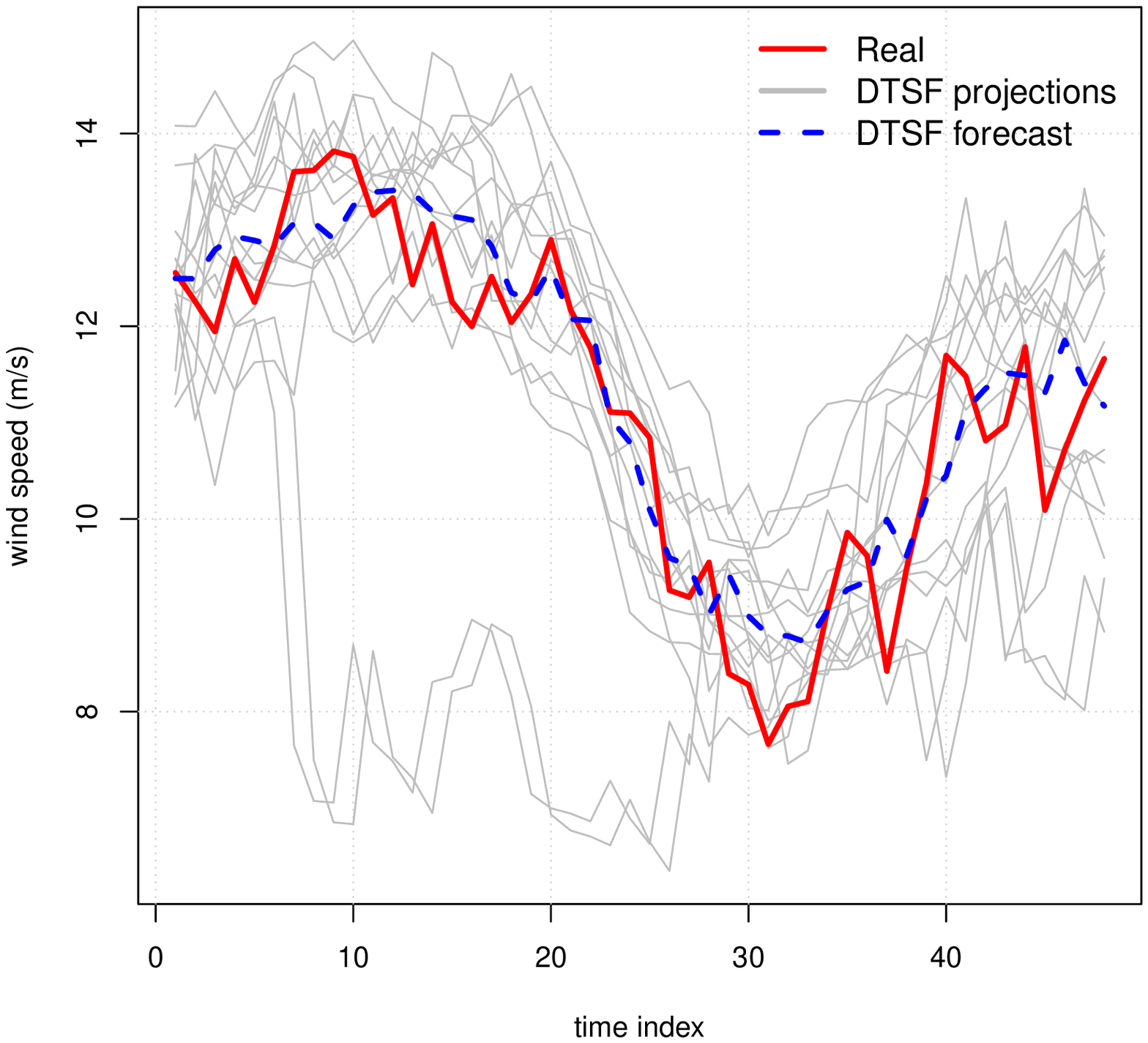}
   }
   \subfigure[Prediction intervals using boxplots.]{
            \label{fig:006}
            \includegraphics[width=0.45\textwidth]{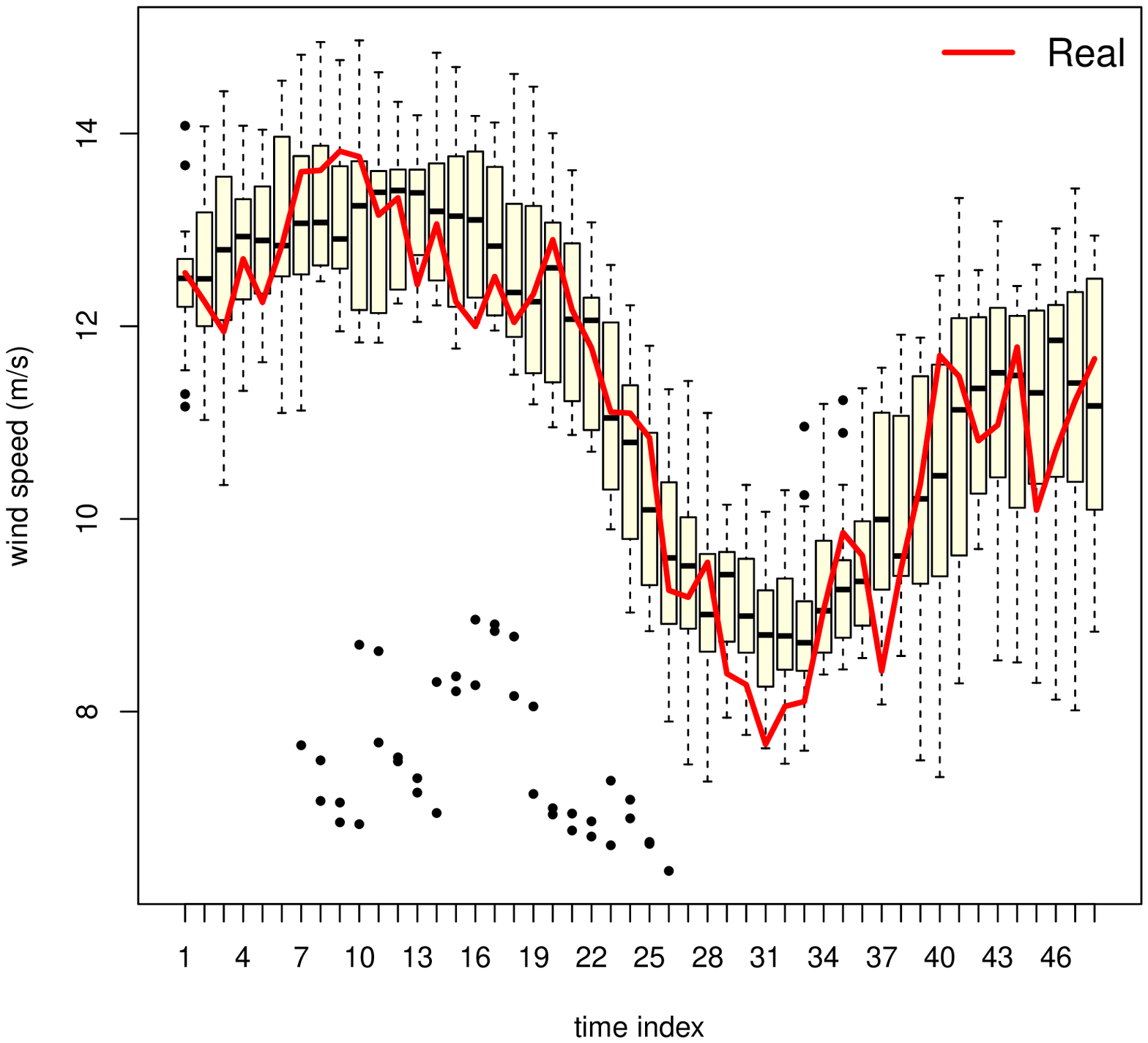}
   }
   \caption{Projected values using DTSF and their respective boxplots used for estimating prediction intervals.}
   \label{fig:007}
\end{figure}

Figure \ref{fig:005} shows the projected values of the 15 best matches using their respective similarity functions for October 26, 2015. The median value of the projections are used as the final forecast. In addition, the real wind speed is shown in Figure \ref{fig:005}. It can be seen that most of the projected values shows a similar pattern as compared to the real wind speed. Furthermore, using a boxplot representation, prediction intervals can be estimated using the interquartile distance, as shown in Figure \ref{fig:006}.

\section{Discussion and conclusion}
\label{sec:conclusion}
As described in the previous section, results indicate that the DTSF is extremely competitive, having achieved, on average, the best performance in the real case wind speed forecasting as compared to standard statistical and machine learning models. Furthermore, the DTSF is very intuitive and based on elementary statistical procedures, such and the scan statistics and the linear regression. One limitations of the DTSF is the requirement of a large database. For smaller time series, standard statistical models such as exponential smoothing and arima models are preferable.

The DTSF can also be applied to qualitative analysis. One may investigate the causes of similar patterns in a time series. Future work aims at combining DTSF, statistical and machine learning models in order to further improve prediction, as suggested by \citet{makridakis2018statistical}. Ongoing work aims at comparing the DTSF performance using multiple time series data.

\section*{Acknowledgements}

The authors thank CEMIG and CNPq for financial support, project numbers PQ-308361/2014-8, CNPq-402070/2016-0 and APQ-03813-12/GT555.






\bibliographystyle{abbrv} 
\bibliography{sample}







\end{document}